\documentclass{article}

\usepackage{epsfig, graphics}
\usepackage{latexsym,subfigure}
\usepackage{amsmath,amssymb,amsthm,amsfonts}
\usepackage[numbers]{natbib}
\usepackage{algorithm,algorithmic,bm, color}

\title{Real Time Bid Optimization with Smooth Budget Delivery in Online Advertising}

\author{Kuang-Chih Lee\\klee@turn.com\\Turn Inc\\
\and
Ali Jalali\\ajalali@turn.com\\Turn Inc\\
\and
Ali Dasdan\\adasdan@turn.com\\Turn Inc}

\begin{document}
\maketitle

\begin{abstract}
Today, billions of display ad impressions are purchased on a daily
basis through a public auction hosted by real time bidding (RTB)
exchanges. A decision has to be made for advertisers to submit
a bid for each selected RTB ad request in milliseconds. Restricted by
the budget, the goal is to buy a set of ad impressions to reach as
many targeted users as possible. A desired action (conversion),
advertiser specific, includes purchasing a product, filling out a
form, signing up for emails, etc. In addition, advertisers typically
prefer to spend their budget smoothly over the time in order to reach
a wider range of audience accessible throughout a day and have a
sustainable impact. However, since the conversions occur rarely and
the occurrence feedback is normally delayed, it is very challenging to
achieve both budget and performance goals at the same time. In this
paper, we present an online approach to the smooth budget delivery
while optimizing for the conversion performance. Our algorithm tries
to select high quality impressions and adjust the bid price based on
the prior performance distribution in an adaptive manner by
distributing the budget optimally across time. Our experimental
results from real advertising campaigns demonstrate the effectiveness
of our proposed approach.
\end{abstract}

\section{Introduction}
\label{sc:introduction}
In recent years, the amount of ad impressions sold through real time
bidding (RTB) exchanges has had a tremendous growth. RTB exchanges
provide a technology for advertisers to algorithmically place a bid on
any individual impression through a public auction. This functionality
enables advertisers to buy inventory in a cost effective manner, and
serve ads to the \emph{right} person in the \emph{right} context at
the \emph{right} time. However, in order to realize such
functionality, advertisers need to intelligently evaluate each
impression in real time. Demand-side platforms (DSPs) offer such a
solution called real time bid optimization
\cite{KLee:EstimateCVRTurn,CRerlich:BidOptimization} to help
advertisers find the optimal bid value for each ad request in
milliseconds close to a million times per second.

The process of real time bid optimization tries to maximize the
campaign performance goal under the delivery constraint within the
budget schedule. The performance goals typically can be specified by
minimizing cost-per-click (CPC) or cost-per-action (CPA), as well as
by maximizing click-through-rate (CTR) or action-rate (AR). Typically,
a smooth budget delivery constraint, expressed as not buying more than
a set fraction of the impressions of interest before a set time, is
used to prevent the campaign from finishing the budget prematurely or
avoiding a bursty spending rate. This constraint generally helps the
advertisers to have sustainable influence with their ads, avoid
pushing large amount of ads in peak traffic (while performance may be
degraded), and explore a broader range of audience.

\begin{figure*}[t]
\centering
\subfigure[Premature Stop]{
    \includegraphics[width=1.8in]{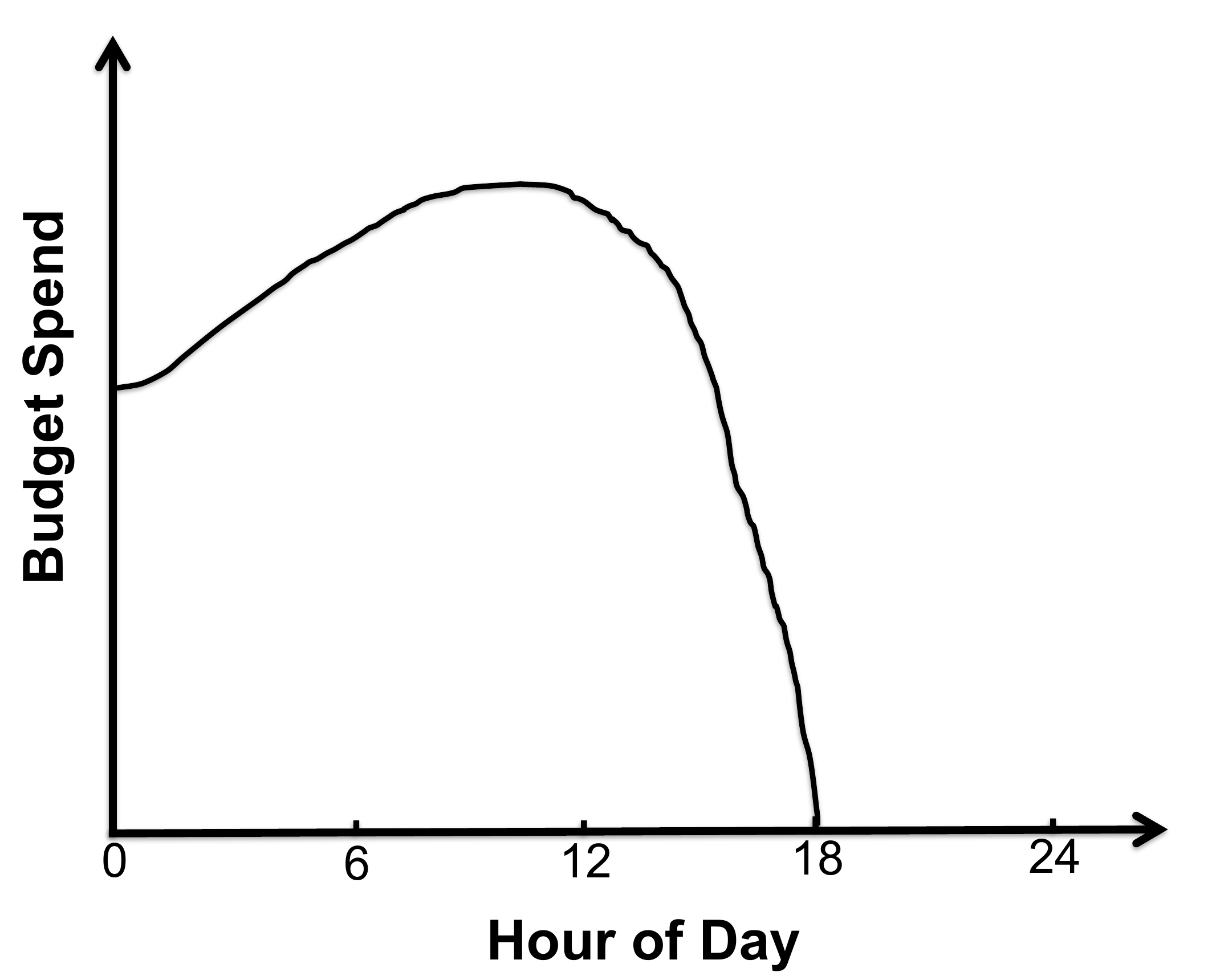}
    \label{fig:premature-budget}
}
\subfigure[Fluctuating Budget]{
    \includegraphics[width=1.8in]{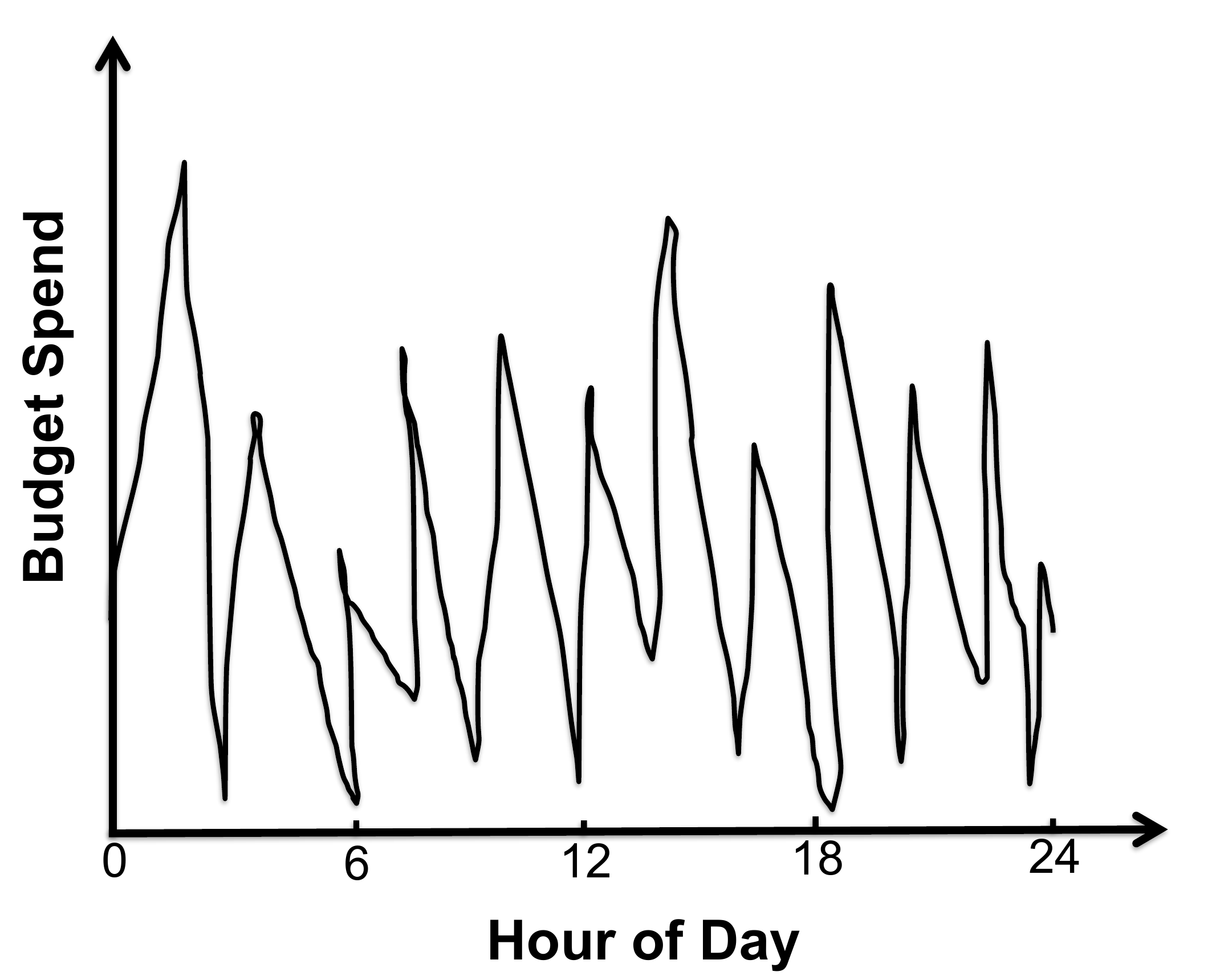}
    \label{fig:fluctuating-budget}
}
\subfigure[Uniform Pacing]{
    \includegraphics[width=1.8in]{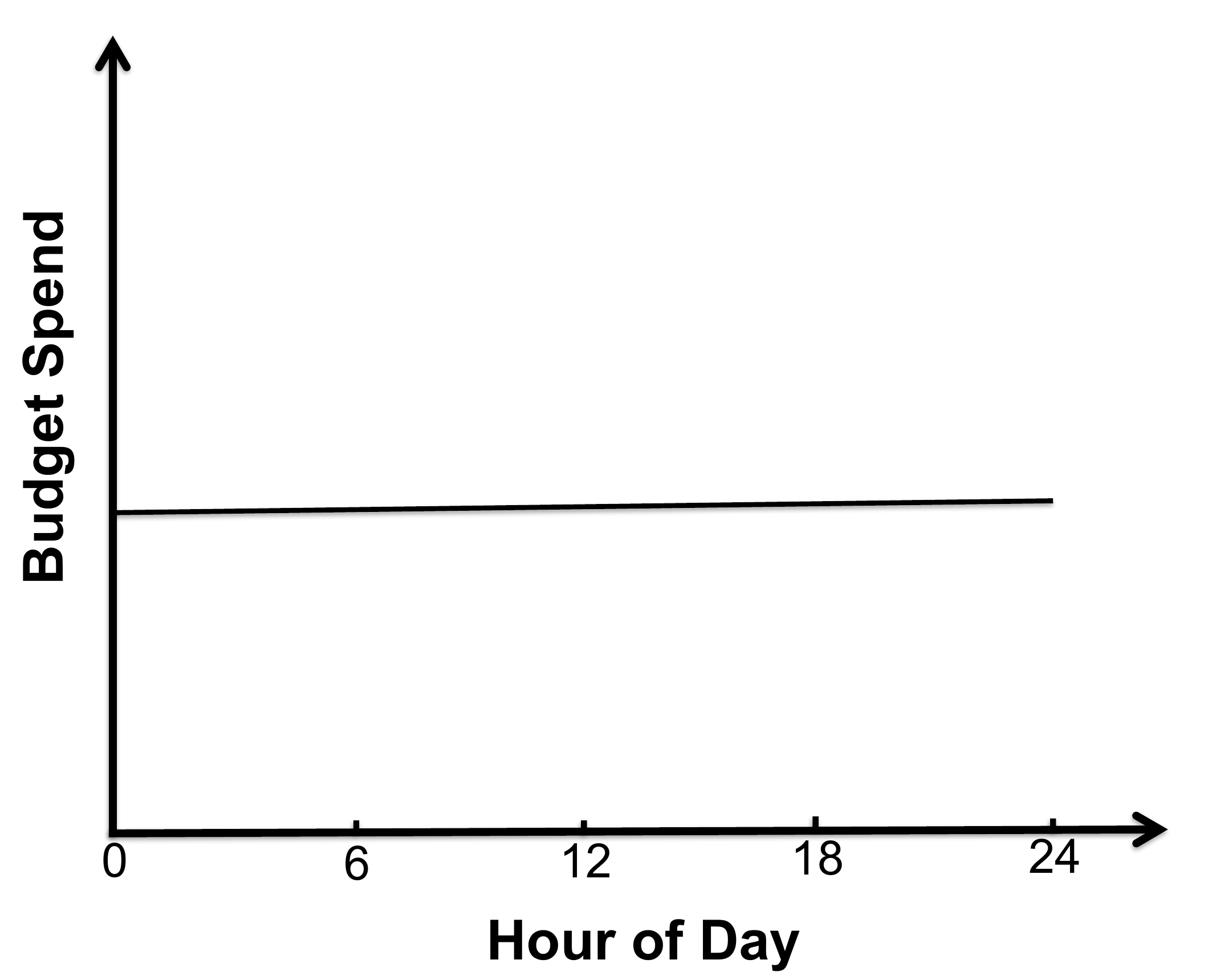}
    \label{fig:uniform-budget}
}

\subfigure[Traffic Based Pacing]{
    \includegraphics[width=1.8in]{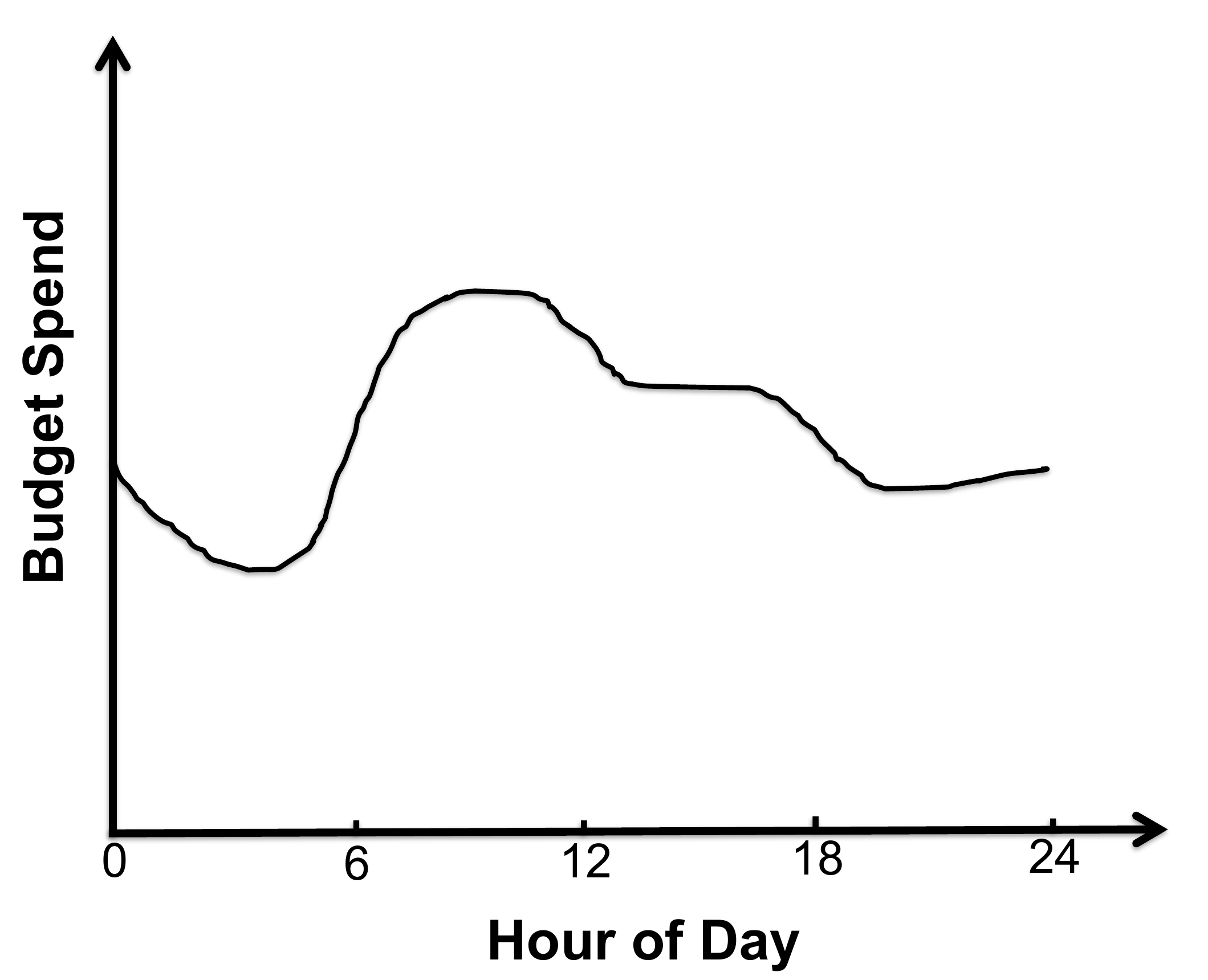}
    \label{fig:traffic-budget}
}
\subfigure[Performance Based Pacing]{
    \includegraphics[width=1.8in]{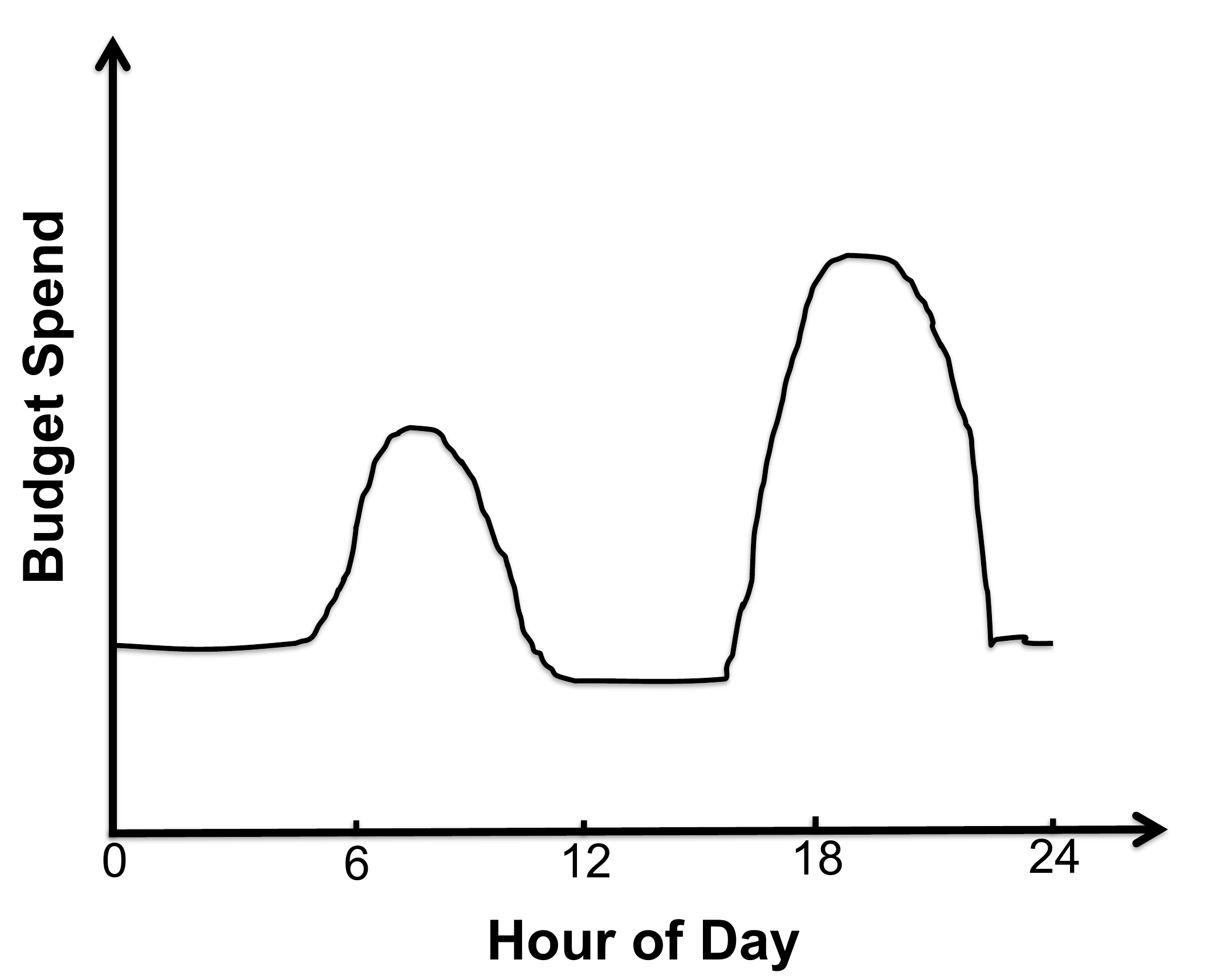}
    \label{fig:performance-budget}
}
\caption{Illustration of different budget pacing schemes with respect
  to the portion of the budget spent every time interval. Integrating
  these plots gives the portion of the budget spend from the beginning
  of the day.}
\label{fig:BudgetPacing}
\end{figure*}

It is challenging to perform real time bid optimization in a RTB
environment for many reasons, including the following. Firstly, the
decision of placing a bid and evaluation of the bid price needs to be
performed per ad request in few milliseconds. In addition, top DSPs
typically receive as many as a million ad requests per second while
hundreds of millions of users simultaneously explore the web around
the globe. The short latency and high throughput requirements
introduce extreme time sensitivity on the process. Secondly, lots of
information is missing in the real time evaluation of the individual
ad requests, e.g., the feedback on previous decisions has normally a
long delay in practice. More specifically, the collection of click
information is delayed because of the duplication removal during the
logging process. On the other hand, most of the view-through actions
often require up to seven days to be converted and attributed to the
corresponding impressions. Finally, click and conversion events are
usually very rare for non-search advertisement and therefore the
variance will be large while estimating the past performance metrics.

In this paper, we present an online approach to optimize the performance 
metrics while satisfying the smooth delivery constraint for
each campaign. Our approach first applies a control feedback loop to
iteratively estimate the future spending rate in order to impose
smooth delivery constraints. Then, the spending rate is used to select
high quality impressions and adjust the bid price based on the prior
performance distribution to maximize the performance goal.

The rest of the paper is organized as follows. In
\S~\ref{sc:background}, we formulate our problem and detail previous
related work. In \S~\ref{sc:bidopt}, we describe our proposed approach
of online bid optimization. Various practical issues encountered
during bid optimization and the proposed solutions are discussed in
\S~\ref{sc:issues}. Thorough experimental results are presented in
\S~\ref{sc:exp}, and in \S~\ref{sc:conclusion} we conclude by a
discussion of our approach and possible future work.

\section{Background and Related Work}
\label{sc:background}
In this section, we first formulate the problem of bid optimization as
an online linear programming, and then discuss the previous related
work in the literature, and explain why those proposed solutions are
not suitable for our online bid optimization problem in practice.

\subsection{Problem Setup}
\label{sc:problem}
Let us consider the online bid optimization in the following settings:
There are $n$ ad requests arriving sequentially ordered by an index
$i$. An individual advertiser would like to make a decision
represented by an indicator variable $x_i \in \{0,1\}$ for all
$i\in\{1,\ldots,n\}$ whether to place a bid on the $i^{th}$ ad request
or not. We consider a total daily budget $B\in\mathbb{R}$ as the total
cost of acquiring ad inventory. Typically, advertisers would like to
have smooth budget delivery constraint, expressed as not buying more
than a set fraction of the impressions of interest before a set time,
in place to ensure the following two situations will never occur:
\begin{itemize}
\item {\bf Premature Campaign Stop}: Advertisers do not want their
  campaigns to run out of the budget prematurely in the day so as not
  to miss opportunities for the rest of day. Such premature budget
  spend is shown in Fig.~\ref{fig:premature-budget} finishing the
  budget 6 hours early.
\item {\bf Fluctuation in Spend}: Advertisers would like to be able to
  analyze their campaigns regularly and high fluctuations in the
  budget makes the consistency of the results questionable. That is
  why a budget pacing scheme similar to what is shown in
  Fig.~\ref{fig:fluctuating-budget} is not suitable.
\end{itemize}

A simple, yet widely used, budget pacing scheme that meets the smooth
delivery constraints is \emph{uniform pacing} or \emph{even pacing}
shown in Fig.~\ref{fig:uniform-budget}. In this scheme the budget is
uniformly split across the day. There are two main issues with this
simple scheme as follows:
\begin{itemize}
\item {\bf Traffic Issue}: Depending on the target audience, the
  volume of the online traffic varies a lot throughout the day. It
  might be the case that during the first half of the day, we receive
  more relevant traffic comparing to the second half of the day;
  however, uniform budget pacing scheme does not allocate the budget
  accordingly. As a result, either we might not be able to deliver the
  budget by the end of the day or we might be forced to buy low
  quality impressions in the second half of the day. A uniform budget
  pacing with respect to the traffic (as opposed to with respect to
  the time) might resolve this issue to some extent. Such scheme is
  depicted in Fig.~\ref{fig:traffic-budget}.
\item{\bf Performance Issue}: The quality of the online traffic
  changes over the course of the day for different groups of
  audience. Whether this quality is being measured by CPC, CPA, CTR or
  AR, the budget pacing algorithm should allocate most of the budget
  to time periods of the day with high quality. Such scheme is
  depicted in Fig.~\ref{fig:performance-budget} and often has few
  picks for the periods with high quality. This potentially can cause
  high fluctuations that might violate smooth delivery constraints.
\end{itemize}
Balancing the traffic and performance under smooth delivery
constraints is challenging. In this paper, we propose a scheme that
resolves both of these issues simultaneously.

In order to enforce the smooth delivery constraints (explained further
in \S~\ref{sc:introduction}), the overall daily budget $B$ can be
broken down into a sequence of time slot schedules $\{b_1, \ldots,
b_T\}$, where, $b_t\in\mathbb{R}$ represents the allocated budget to
the time slot $t$, and $\sum_{t=1}^T\, b_t = B$. In the next section,
we will introduce how to impose different pacing strategies to assign
$b_t$'s in order to select higher quality impressions. Each ad request
$i$ is associated with a value $v_i\in\mathbb{R}$ and a cost
$c_i\in\mathbb{R}$. The value $v_i$ represents the true value for the
advertiser if the given ad request $i$ has been seen by an
audience. The cost $c_i$ represents the actual advertiser cost for the
ad request $i$ paid to the publisher serving the corresponding
impression. In summary, the bid optimization problem with smooth
budget delivery constraint can be formulated as
\begin{align}
\label{eq:objective}
\mbox{\bf maximize} \;\;& \;\;\sum^n_{i=1} v_ix_i \notag \\
\mbox{\bf subject to} \;\;& \;\;\sum_{j\in\mathbb{I}_t} c_jx_j \leq b_t\qquad\forall t\in\{1,\ldots,T\}, 
\end{align}
where, $\mathbb{I}_t$ represents the index set of all ad requests
coming in the time slot $t$. Obviously this optimization problem is an
offline formulation due to the fact that the cost and value of future
ad requests are not clear at the time of decision on $x_i$. More
precisely, after the (current) incoming ad request $i$ is received,
the online algorithm of bid optimization must make the decision $x_i$
without observing further data. For dynamic bidding campaigns, the
optimization process also needs to estimate $\widehat{c}_i$ as the bid
price at the same time. Please note that the bid price is not
equivalent to the cost $c_i$ for the incoming ad request $i$, because
the cost is determined by a second price auction in the RTB
exchange. More clearly, one should bid $\widehat{c}_i = c_i +
\epsilon_i$ to be able to win the second price auction and actually
pay $c_i$. The value of $\epsilon_i$ is determined based on the
auction properties and is unknown to the bidder at the bidding time.

\subsection{Related Work}
Eq.~\eqref{eq:objective} is typically called online linear
programming, and many practical problems, such as online bidding
\cite{CBorgs:BidOptAuctions,WZhang:JointOptimizationBidBudget}, online
keyword matching \cite{YZhou:BiddingKnapsack}, online packing
\cite{JFeldman:OnlineStochasticPacking}, and online resource
allocation \cite{VMirrokni:OnlineSmoothDelivery}, can be formulated in
the similar form. However, we do not attempt to provide a
comprehensive survey of all the related methods as this has been in a
number of papers
\cite{SAgrawal:OnlineLinearProgramming,MBabaioff:OnlineAuction}. Instead
we summarize couple of representative methods in the following.

Zhou \emph{et al.} \cite{YZhou:BiddingKnapsack} modeled the budget
constrained bidding optimization problem as an online knapsack
problem. They proposed a simple strategy to select high quality ad
requests based on an exponential function with respect to the budget
period. As time goes by, the proposed algorithm will select higher and
higher quality of ad requests. However, this approach has an
underlying assumption of unlimited supply; i.e., there are infinite
amount of ad requests in the RTB environment. This assumption is
impractical especially for those campaigns with strict audience
targeting constraints.

Babaioff \emph{et al.} \cite{MBabaioff:DynamicPricing} formulated the
problem of dynamic bidding price using multi-armed bandit framework,
and then applied the strategy of upper confidence bound to explore the
optimal price of online transactions. This approach does not require
any information about the prior distribution. However, multi-armed
bandit framework typically needs to collect feedback quickly from the
environment in order to update the utility function. Unfortunately,
the collection of bidding and performance information has longer delay
for display advertising in RTB environment.

Agrawal \emph{et al.} \cite{SAgrawal:OnlineLinearProgramming} proposed
an general online linear programming algorithm to solve many practical
online problems. First they applied the standard linear programming
solver to compute the optimal dual solution for the data which have
been seen in the system. Then, the solution for the new instance can
be decided by checking if the dual solution with the new instance
satisfies the constraint. The problem is that the true value $v_i$ and
cost $c_i$ for the incoming ad request is unknown when it arrives to
the system. If $v_i$ and $c_i$ is estimated by some statistical models
or other alternative solutions, the dual solution needs to be
re-computed more frequently for each campaign in order to impose
budget constraints accurately. This introduces high computational cost
in the real time bidding system.

\section{Online Bid Optimization}
\label{sc:bidopt}
In this section, we detail our method of online bid optimization. We
first revisit the smooth delivery constraint and explain how we
control the spending rate adaptively when each ad request comes
sequentially. Afterwards, we discuss how we can iteratively apply the
spending information to select ad requests and adjust their bid price
to optimize the objective function.

One should recognize two different classes of campaigns: (i) Flat CPM
campaigns, and, (ii) Dynamic CPM (dCPM) campaigns. The main difference
between the two types is that the first one submits a flat bid price
whereas the second one optimized the bid price. Both also need to
decide whether to bid on an ad request. The metric for the goodness of
the decision with flat CPM campaigns is typically either CTR or AR;
while the metric for dCPM is typically effective CPC (eCPC) or
effective CPA (eCPA). The difference between CPA and eCPA is that CPA
is the goal to reach while eCPA is what is actually
realized. Regardless of the type of the campaign, we try to optimize
the following goal:
\begin{equation}
\begin{aligned}
&\min\;\;\;\; \text{-CTR, -AR, eCPC or
    eCPA}\\ &\;\text{s.t.}\;\;\;\;|\sum_t^T\text{\bf s}(t) -
  B|\leq\epsilon\\ &\qquad\;\;\left|\text{s}(t) -
  b_t\right|\leq\delta_t\qquad\qquad\forall t\in\{1,\ldots,T\}\\ &\qquad\;\; eCPM\leq M
\end{aligned}
\end{equation}
where the first constraint is the total daily budget constraint (where
$s(t)$ is the budget spent at time slot $t$), the second constraint
enforces smooth delivery according to the schedule $b_t$ and the third
constraint requires that eCPM does not exceed the cap $M$. The last
constraint makes a dCPM campaign appear like a CPM campaign in average
over time, hence, the use of CPM in dCPM.

In this formulation, the optimization parameter is $b_t$, since the
total budget $B$ and average impression cost cap $M$ are set by the
advertiser. We detail our budget pacing scheme in the rest of this
section and show how we improve this optimization by smart budget
pacing.

\subsection{Smooth Delivery of Budget}
The original idea of budget pacing control is to take the daily budget
as input and calculate a delivery schedule in real-time for each
campaign. Based on the delivery schedule, the DSP will try to spread
out the actions of acquiring impressions for each campaign throughout
the day. We break down a day into $T$ time slots and in each time
slot, we assign a budget to be spent by each campaign.

In the time slot $t$, the spend of acquiring inventory is considered
to be proportional to the number of impressions served at that time
slot; assuming the price of individual impressions for a particular
campaign remains approximately constant during that time slot. In
reality, the length of the time slot should be chosen such that the
variance of individual impression price for each campaign is
small. Our analysis shows that this assumption holds on our real data
if the length of the time slot is properly chosen.

To each campaign, we assign a \emph{pacing rate} for each time slot
$t$. The pacing rate is defined to be the portion of incoming ad
requests that this campaign would like to bid on. To see the
relationship of the pacing rate with other parameters of our proposed
bid optimization system, consider the following equation derived from
constraints in Eq.~\ref{eq:objective}:

\begin{align}
\label{eq:OriginalProportion}
\mbox{\bf s}(t) &=  \sum_{j\in\mathbb{I}_t} c_jx_j \notag \propto \mbox{\bf imps}(t) \notag \\
&\propto\mbox{\bf reqs}(t) \frac{\mbox{\bf bids}(t)}{\mbox{\bf reqs}(t)}\frac{\mbox{\bf imps}(t)}{\mbox{\bf bids}(t)}  \\ 
&\propto\mbox{\bf reqs}(t) \cdot \mbox{\bf pacing\_rate}(t) \cdot \mbox{\bf win\_rate}(t) \notag
\end{align}
 
Here, {\bf s}$(t)$ is the dollar amount of money spent, $\text{\bf
  reqs}(t)$ is the number of incoming ad requests that satisfy the
audience targeting constraints of the campaign, $\text{\bf bids}(t)$
is the number of ad requests that this campaign has bid on, and
finally $\text{\bf imps}(t)$ is the total number of impressions of the
campaign, i.e., the bids that are won in the public auction, all
during the time slot $t$. With these definitions, we naturally define
the ratio of bids to ad requests as \emph{pacing rate} and the ratio
of impressions to bids as \emph{win rate}. Notice that if we assume
those ad requests that satisfy the audience targeting of a certain
campaign are uniformly distributed across all of the incoming ad
requests, then, one can replace $\text{\bf reqs}(t)$ with some
constant times the total number of incoming ad requests. That constant
can be absorbed by the proportion in \eqref{eq:OriginalProportion}.

To make progress, we want to take a dynamic sequential approach in
which, we get a feedback from the previous time slot spend and adjust
our pacing rate for the next time slot. By working on the proportion
\eqref{eq:OriginalProportion}, the pacing rate for the next time slot
$t+1$ can be obtained by a simple recursive equation as follows:
\begin{align}
\label{eq:OriginalRecursion}
&\mbox{\bf pacing\_rate}(t\!+\!1) \\
&= \mbox{\bf pacing\_rate}(t)\frac{\mbox{\bf s}(t\!+\!1)}{\mbox{\bf s}(t)}
    \frac{\mbox{\bf reqs}(t)}{\mbox{\bf reqs}(t\!+\!1)} \frac{\mbox{\bf win\_rate}(t)}{\mbox{\bf win\_rate}(t\!+\!1)} \notag \\
&= \mbox{\bf pacing\_rate}(t)\frac{b_{t\!+\!1}}{\mbox{\bf s}(t)} 
\frac{\mbox{\bf reqs}(t)}{\mbox{\bf reqs}(t\!+\!1)} \frac{\mbox{\bf win\_rate}(t)}{\mbox{\bf win\_rate}(t\!+\!1)}\notag
\end{align}
where, $\text{\bf reqs}(t\!+\!1)$ and {\bf win\_rate}$(t\!+\!1)$
represent the predicted number of ad requests and the predicted
winning rate for the bids in the next time slot $t+1$. One can do such
predictions using historical data keeping in mind that we are only
interested in the ratio of these parameters to their previous values
and not necessarily their absolute values. This recursive definition
introduces a simple adaptive feedback control for smooth budget
pacing.

The future spend, i.e., $\text{\bf s}(t\!+\!1)$, in
\eqref{eq:OriginalRecursion} is set to be equivalent to the ideal
desired spend $b_{t\!+\!1}$ at time slot $t+1$ in order to impose the
budget constraint. Different choices of $b_{t\!+\!1}$ introduces
different strategies for the budget pacing. For example, one simple
strategy, called uniform pacing, tries to spend the budget evenly for
the given campaign throughout the day. This strategy can be easily
implemented by defining future spend $b_{t\!+\!1}^{u}$ ($u$ to denote
uniform) as
\begin{align}
b^{u}_{t\!+\!1} \!=\! \Big(B\!-\!\sum_{m=1}^t\text{\bf s}(m)\Big)\frac{L(t+1)}{\sum_{m=t+1}^T\; L(m)}
\label{eq:UniformGeneral}
\end{align}
where, the first factor represents the remaining budget of the day and
the second factor is the ratio of the length of the time slot $t+1$ to
the remaining time in the day; i.e., $L(t)$ is the length of the time
slot $t$. If time slots have equal length, one can simplify
\eqref{eq:UniformGeneral} to get
\begin{align}
b^{u}_{t\!+\!1} = \Big(B-\sum_{m=1}^t{\bf s}(m)\Big)\frac{1}{T-t}.
\label{eq:UniformSimple}
\end{align}

Uniform pacing is not necessarily the best strategy as discussed in
\S~\ref{sc:problem}. We propose a strategy to spend more money on the
time slots where a particular campaign has more chance to get events
of interest (clicks or conversions). To do so, we look at the campaign
history data and measure the performance of the campaign during each
time slot. Based on this measurement, we build a discrete probability
density function described by a list of click or conversion
probabilities: $p_0, \ldots, p_T$ assuming $T$ time-slots per day, and
$\sum_t^T p_t = 1$. Now at each time slot, we compute the ideal
spending $b^{p}_{t\!+\!1}$ ($p$ to denote probabilistic) for the
next time slot as
\begin{align}
b^{p}_{t\!+\!1} \!=\! \Big(B\!-\!\sum_{m=1}^t\text{\bf s}(m)\Big)\frac{p_{t\!+\!1}\cdot L(t\!+\!1)}{\sum^T_{m=t\!+\!1}p_m\cdot L(m)}.
\label{eq:PerformanceGeneral}
\end{align}
Similar to the uniform pacing case, if the time slots have equal lengths, this can be simplified to
\begin{align}
b^{p}_{t\!+\!1} \!=\! \Big(B\!-\!\sum_{m=1}^t\text{\bf s}(m)\Big)\frac{p_{t\!+\!1}}{\sum^T_{m=t\!+\!1}p_m}.
\label{eq:PerformanceSimple}
\end{align}

In practice, it is important to notice that if $p_j = 0$ for some $j$,
then that campaign will never spend money during that time slot and
hence, it will never explore the opportunities coming up during time
slot $j$. To prevent this situation, one can split the budget and use
a combination of the above two strategies. This way, there will be
always a chance to explore all possible opportunities.

After the pacing rate is calculated, each campaign can apply this
information to adaptively select certain portions of high quality
impressions, as well as adjust the bid price in order to maximize the
objective function. We explain those details in the next subsections.

\subsection{Selection of High Quality Ad Requests - Flat CPM Campaigns}
We consider two cases of flat CPM and dynamic CPM separately as the
former, unlike the latter, does not need a bid price calculation. For
flat CPM campaigns that always submit a fixed bid price $c^*$ to RTB
exchanges, the goal is to simply select a set of ad requests to bid on
considering the current time slot pacing rate. Since we do not know if
the current incoming ad request will eventually cause a click
or conversion event during the time of bid optimization, the true value
of the ad request is estimated by the prediction of its CTR or AR
using a statistical model. The details of the offline training process
of CTR or AR prediction is described in \S~\ref{sc:estimatecvr}.

\begin{figure}[t]
  \centering
  \includegraphics[width=2.0in]{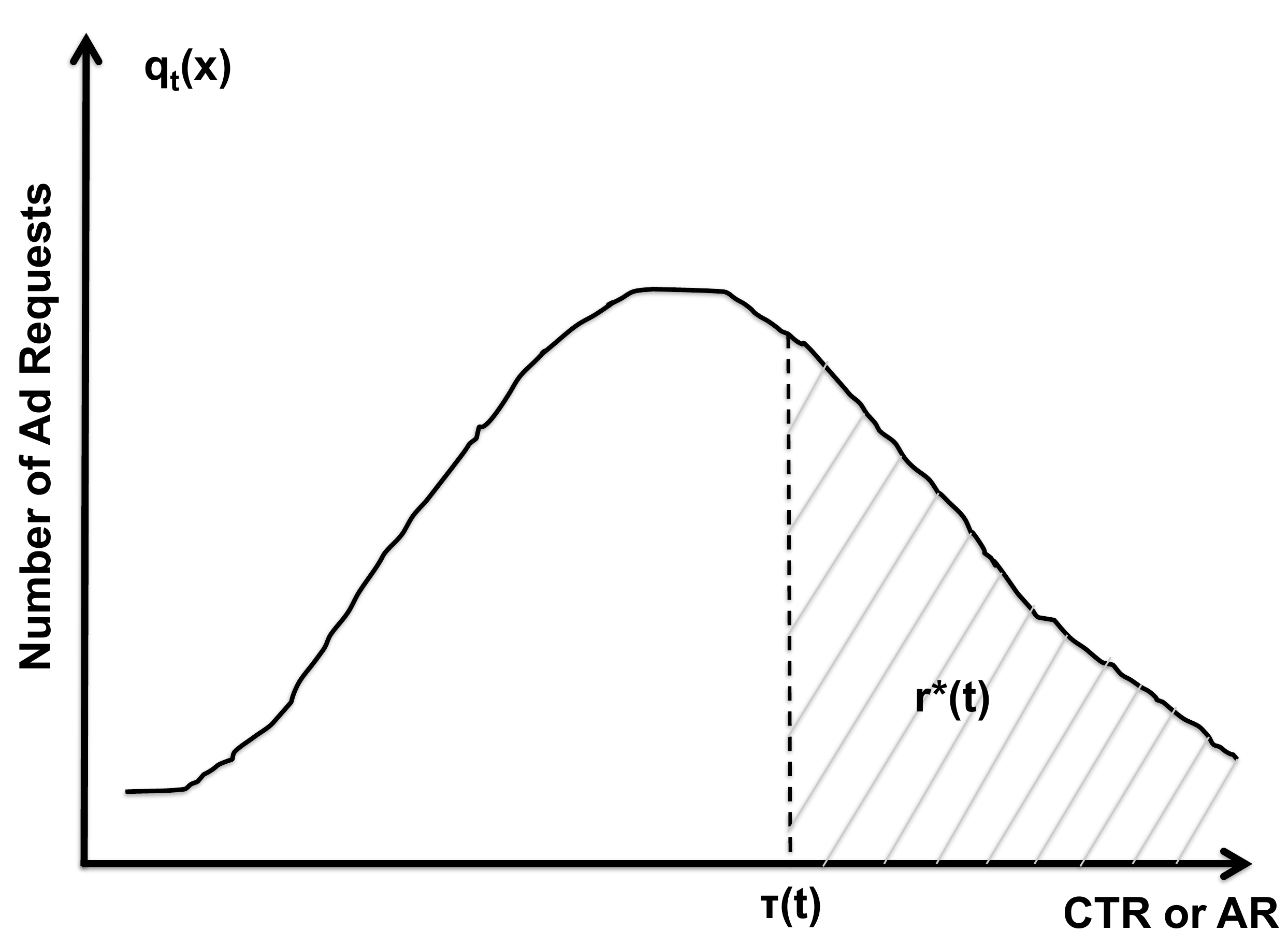}
  \caption{Illustration of our performance based budget pacing for flat CPM campaigns.}
   \label{fig:flatCPMhistogram}
\end{figure}

Notice that to fulfill the smooth delivery constraint, we require a
minimum number of impressions given by $$\text{\bf imps}^*(t) =
\frac{\text{\bf s}(t)}{c^*}.$$ This number of impressions can be
reached only if we have $$\text{\bf bids}^*(t) =\frac{\text{\bf imps}^*(t)}{\text{\bf
    win\_rate}(t)}.$$ Similarly, to get these
many bidding opportunities, we expect to have $$\text{\bf reqs}^*(t)
=\frac{\text{\bf bids}^*(t)}{\text{\bf pacing\_rate}(t)}.$$ Now, we
are going to select these ad requests from the set of incoming ad
requests whose chance of a click or a conversion is high. To do so, we
construct an empirical histogram of CTR or AR distribution $q_t(x)$
based on the historical data for each campaign, where $q_t(x)$
represents the number of ad requests in time slot $t$ that are
believed to have CTR or AR of $x$, e.g., see
Fig.~\ref{fig:flatCPMhistogram}. Our online algorithm finds a
threshold $\tau(t)$ in the time slot $t$ to filter ad requests in the
region of $q_t(x)$ that has low CTR or AR rate such that the smooth
delivery is fulfilled. Such threshold can be formulated as
\begin{align}
\tau(t) = \arg\min_{x} \left| \int^1_{x} q_t(s)ds - \text{\bf reqs}^*(t) \right|
\label{eq:ThresholdflatCPM}
\end{align}

In practice, since the CTR or AR distribution $q_t(x)$ is not computed
frequently, it introduces some oscillations for the threshold
$\tau(t)$ in different time slots if $q_t(x)$ is not close to the
current ad request distribution. Note that this mismatch is very
probable since $q_t(x)$ is generated from historical data that might
not perfectly correlate with the current reality. In order to prevent
this situation, we evaluate a confidence interval of the threshold
parameter $\tau(t)$. First, we incrementally update the mean
$\mu_{\tau}(t)$ and variance $\sigma_{\tau}(t)$ of the threshold
$\tau(t)$ using the online adaptation as follows
\begin{align}
&\mu_{\tau}(t) = \mu_{\tau}(t\!-\!1) + \frac{1}{t}\left( \tau(t)\! -\! \mu_{\tau}(t-1)\right)\\
&\sigma^2_{\tau}(t) = \frac{t\!-\!1}{t}\sigma^2_{\tau}(t\!-\!1) \!+\! \frac{1}{t}\left(\tau(t)\! -\! \mu_{\tau}(t\!-\!1) \right)\left(\tau(t) \!-\! \mu_{\tau}(t) \right)\notag
\end{align}
Second, assuming that $\tau(t)$ comes from a Gaussian distribution, we
bound $\tau(t)$. The upper and lower bounds of the threshold $\tau(t)$
can be stated as $\mu_{\tau}(t)+
\gamma\frac{\sigma_{\tau}(t)}{\sqrt{d}}$ and $\mu_{\tau}(t)-
\gamma\frac{\sigma_{\tau}(t)}{\sqrt{d}}$, respectively, where $d$ is
the number of days we looked into the history of the data to make the
statistics. The critical value $\gamma=1.96$ provides $95\%$
confidence interval. The upper bound and lower bound of the CTR or AR
threshold are updated in each time slot.

Putting all together, when an ad request comes to the system, its CTR
or AR value is first estimated by the statistical model. If the
predicted value is larger than the upper bound of the threshold, this
ad request will be kept and the fixed bid price $v^*$ will be
submitted to the RTB exchange. If the predicted value is smaller than
the lower bound of the threshold, this ad request will be simply
dropped without further processing. If the predicted value is in
between the upper and lower bounds, the ad request will be selected at
random with probability equal to {\bf pacing\_rate}$(t)$. This scheme,
although approximate, ensures that the smooth delivery constraint is
met while the opportunity exploration continues on the boundary of
high and low quality ad requests.

\subsection{Selection of High Quality Ad Requests - Dynamic CPM Campaigns}
\label{sc:adjustbidprice}

For dCPM campaigns, which are free to change the bid price $c_i$
dynamically for each incoming ad request, the goal is to win enough
number of high quality impressions for less cost. We first construct
bidding histogram, with $c^*$ being the historical average of $c_i$ in
time slot $t$, to represent the statistics of good and bad impressions
as discussed in the previous subsection. Then, starting from a base
bid price, we scale the bid price up or down properly considering the
current $\text{\bf pacing\_rate}(t)$ to meet the budget obligation. We
explain the second step in this subsection. For simplicity and without
loss of generality, we will focus on CPA campaigns.

Notice that $\text{\bf pacing\_rate}(t)$ controls the frequency of
bidding; however, if the submitted bid price is not high enough, the
campaign might not win the impression in the public auction. On the
other hand, if the bid price is very high, then cost per action might
rise (even in second price auction as the other bidders will increase
their bid price too). To adjust the bid price, defining thresholds
$0\leq\beta_1\leq\beta_2\leq 1$, we consider three regions for the
pacing rate: (a) safe region: when $\text{\bf
  pacing\_rate}(t)\leq\beta_1$ and there is no under delivery issue
due to audience targeting, (b) critical region: when
$\beta_1\leq\text{\bf pacing\_rate}(t)\leq\beta_2$ and the delivery is
normal, and (c) danger region: when $\beta_2\leq\text{\bf
  pacing\_rate}(t)$ and the campaign has a hard time to find and win
enough impressions. We treat each of these cases separately.

Typically dCPM campaigns work towards meeting or beating a CPA goal
$G$ (compared with eCPA). We use this goal value to define a base bid
price $u_i = \text{AR}\times\text{G}$, where AR is the predicted AR
for the current ad request. We discuss the estimation of the AR in the
next subsection. If our campaign is in the critical region, we submit
$\widehat{c}_i=u_i$ as our bid price since the campaign is doing just
fine in meeting all the obligations.

For the case where our campaign is in the safe region, we start
learning the best bid price from the second price auction. In
particular, we consider the difference between our submitted bid price
and the second price we actually pay for both good and bad
impressions. If the AR estimation algorithm is a high quality
classifier, one expects to see bigger differences for high quality
impressions compared to low quality ones. The reason is that a good
classifier typically generates high AR for high quality impressions
resulting in high values of $u_i$ and hence a bigger difference from
the second price unless all the bidders use the same or similar
algorithm.

Suppose in the past we have submitted $\widehat{c}_i$ as our bid price
and we actually paid $c_i$. For those impressions that led to an
action, we can build the histogram of
$\theta=\frac{c_i}{\widehat{c}_i}$ and find the $\theta^*$ to be the
bottom $1$ or $2$ percentile on that histogram. We propose to submit
$\widehat{c}_i=\theta^*u_i$ as the bid price in this case. Obviously,
this scheme hurts the spending while improving the performance;
however, this is not a problem because the campaign is in the safe
region.

Finally, for campaigns in the danger region, we need to understand why
those campaigns are in this region. There are two main reasons for
underdelivery in this design: (i) The audience targeting constraints
are too tight and hence, there are not enough incoming ad requests
selected for a bid, and, (ii) the bid price is not high enough to win
the public auction even if we bid very frequently. There is nothing we
can do about the first issue, however, we can fix the second issue by
boosting the bid price.

Consider a bid price cap $C$ which in reality is being set by each RTB
exchange. We would like to boost the bid price with parameter
$\rho^*\geq 1$ so that if pacing rate is greater than $\beta_2$, the
parameter $\rho^*$ increases as pacing rate increases. One suggestion
can be a linear increase as
\begin{align}
\rho^* = 1 + \frac{C/c^*\,-1}{1-\beta_2}\left(\text{\bf pacing\_rate}(t)-\beta_2\right).
\end{align}
At the end we submit $\widehat{c}_i=\rho^*u_i$ as our bid
price. Notice that $c^*$ as defined in the beginning of this
subsection is the average historical value of $c_i$ and it dynamically
(and slowly) changes as the market value of the impressions change.

\subsection{Estimation of CTR and AR }
\label{sc:estimatecvr}

Thus far in the paper, we based our algorithm on the top of the
assumption that we have a good system to accurately estimate click
through rate (CTR) and action or conversion rate (AR). In this
section, we will describe how to do this estimation. Again for
simplicity and without loss of generality, we will focus on AR.

This estimation plays a crucial role in bid optimization system for
many reasons including the followings. Firstly, the estimated AR
provides a quality assessment for each ad request helping to decide on
taking action on that particular ad request. Secondly, the base bid
price is set to be the estimated AR multiplied by the CPA goal, which
directly affects the cost of advertising.

There are many proposals for estimating AR in the literature. Since
conversions are rare events, hierarchical structure of features for
each triplet combination of (user, publisher, advertiser) have been
commonly used to smooth and impute the AR for the leaf nodes that do
not have enough conversion
events~\cite{Agarwal:EstimatingRates2,Agarwal:EstimatingRates1,
  KLee:EstimateCVRTurn, Marlin:CollaborativePredictionNRM,
  Zhang:FastComputation}. On the other hand, there are also some
studies that try to cluster users based on their behaviors and
interests and then estimate AR in each user cluster, e.g., see
\cite{Ahmed:BehaviorTargeting, Cerrato:MerketingSegment,
  CRerlich:BidOptimization}. In addition, many standard machine
learning methods, e.g., logistic regressions
~\cite{KLee:EstimateCVRTurn, Richardson:CTRprediction} and
collaborative filtering~\cite{Menon:ResponsePrediction}, are used to
combine multiple AR estimates from different levels in the hierarchy
or user clusters to produce a final boosted estimate.

We use the methodology introduced in \cite{KLee:EstimateCVRTurn} and
make some improvements on the top of that. In this method, we would
like to find the AR for each triplet combination of (user, publisher,
advertiser) by leveraging the hierarchy structure of their
features. Each actual \emph{creative} (the graphic ad to be shown on
the page) is a leaf in the advertiser tree hierarchy. The hierarchy
starts with the root and continues layer after layer by advertiser
category, advertiser, insertion order, package, line item, ad and
finally creative. Using the historical data, one can assign an AR to
each node in this tree by aggregating total number of impressions and
actions of their children as a raw estimate. Same hierarchy and raw
estimations can be done for publisher and user dimensions.

After constructing all three hierarchies with initial raw estimates,
we employ a smoothing algorithm similar to the one discussed in
\cite{Agarwal:EstimatingRates2,Agarwal:EstimatingRates1} to adjust the
raw estimates on different levels based on the similarity and
closeness of (user, publisher, advertiser) triplet on the hierarchy
trees. Then, for each triplet on the leaves of the trees, we run a
logistic regression over the path from that leaf to the root. This
scheme results in a fairly accurate estimation of AR.

\begin{figure*}[!ht]
  \centering
  \includegraphics[width=4.5in]{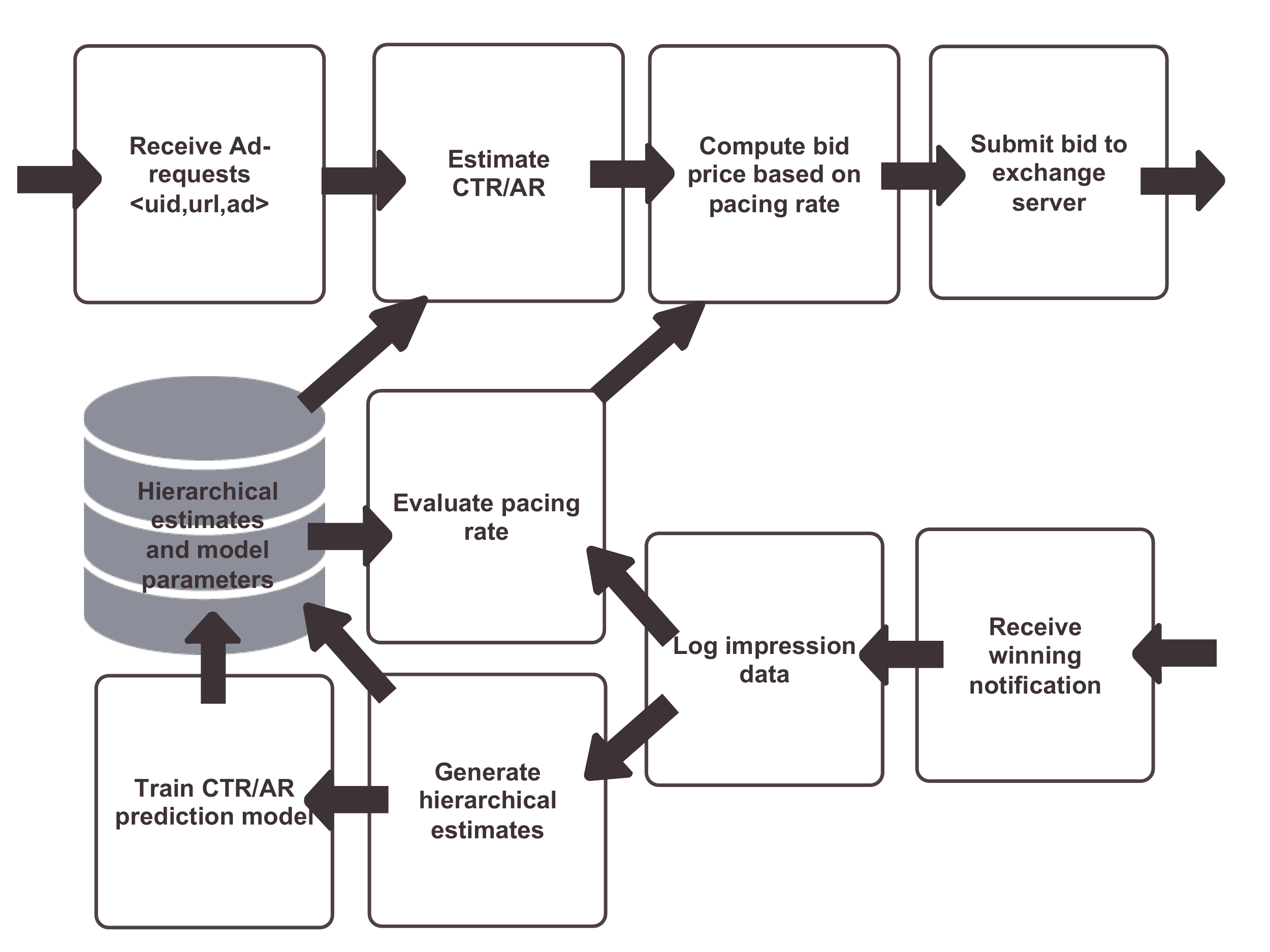}
  \caption{Illustration of ad request workflow in our proposed bid
    optimization framework.}
   \label{fig:data-flow}
\end{figure*}

\section{Practical Issues}
\label{sc:issues}

In this section, we discuss several practical issues encountered
during the implementation of our proposed bid optimization method and
present our current solutions.

\subsection{Cold Start Problem}
When a new campaign started, click or conversion events require some
time to feedback to the system and therefore there is no sufficient
information to perform CTR or AR estimation as well as bid
optimization. This is known as cold start problem and has been well
studied in the literature, e.g., see
\cite{HaibinCheng:MultimediaFeatures,Kanagal:MatrixFactorization,Schein:ColdStart}. We
follow similar ideas to apply content features such as user and
publisher attributes to recommend a list of high quality websites and
audience groups by inferring similarities among existing campaigns. In
addition, a contextual-epsilon-greedy based strategy is performed
during the online bid optimization. If the incoming ad request is
inside one of those recommended publisher or audience groups, a higher
bid is placed. Otherwise, the ad request will be randomly selected
with a default bid price to explore unseen sites and users. As the
campaign gets older and accrues more data, the activity of online
exploration will be decreased and the regular prediction model will
jump to play the major role of bid optimization.

\subsection{Prevention of Overspending}
Since budget spending is controlled by the pacing rate, if there is a
huge amount of ad requests coming all of a sudden, the overall daily
spend might exceed the allocated daily budget $B$. In order to
overcome this problem, several monitoring processes have been
implemented to frequently check the overall daily budget spend as well
as the interval spend in each time slot $t$. If the overall spending
exceeds the daily budget $B$, the campaign will be completely
stopped. If the interval spend exceeds $b_t + \delta$, the bidding
activity will be temporarily paused until the next time slot $t+1$.

\subsection{Distributed Architecture}
Fig.~\ref{fig:data-flow} illustrates the simplified algorithmic flow
chart for each individual ad request. Please note that this workflow
to submit a bid needs to happen within less than 50 milliseconds, and
close to a million ad requests need to be processed in a
second. Therefore, the entire bid optimization is implemented and
parallelized on many distributed computing clusters cross different
data centers. The offline training process utilizes R, Pig, and Hadoop
to generate hierarchical CTR and AR estimates as well as train
campaign-specific prediction models over a large number of
campaigns. The online process streams incoming ad requests to many
servers and evaluates the bid price via a real-time message bus. Our
proposed algorithm works very well in this distributed environment,
and detailed experimental results are presented in the next section.

\begin{table*}[t]
\begin{center}
\begin{tabular}{|c|c|c|c|c|c|c|c|}
\hline
Method & FC1 & FC2 & FC3 & FC4 & FC5 & FC6 & FC7 \\
\hline
Our proposal & 0.0159 &  0.0008 &  0.0003 &  0.0009  &  0.0005 & 0.0008 & 0.0016\\
\hline
Baseline &  0.0041 &  0.0005 &  0.0002 &  0.0003 &  0.0002 & 0.0005 & 0.0007\\
\hline
Improvement (\%) & 286\% & 59\% & 63\% & 155\% & 132\% & 47\% & 124\%\\
\hline
\end{tabular}
\end{center}
\vspace{-0.2in} \caption{CTR improvement for seven selected flat CPM
  campaigns.}
\label{Table:FlatCPM}
\end{table*}

\section{Experimental Results}
\label{sc:exp}

Our proposed framework of bid optimization has been implemented,
tested, and deployed in Turn, a leading DSP in the Internet
advertising industry. In this section, we present simulation results
from our staging environment to compare different strategies of budget
pacing. In addition, we also show results from real campaigns that
serve large amounts of daily impressions in order to demonstrate the
overall performance improvement in terms of CPC or CPA performance
metrics.

\subsection{Comparison of Pacing Strategies }
\begin{figure}[t]
  \centering
  \includegraphics[width=3in]{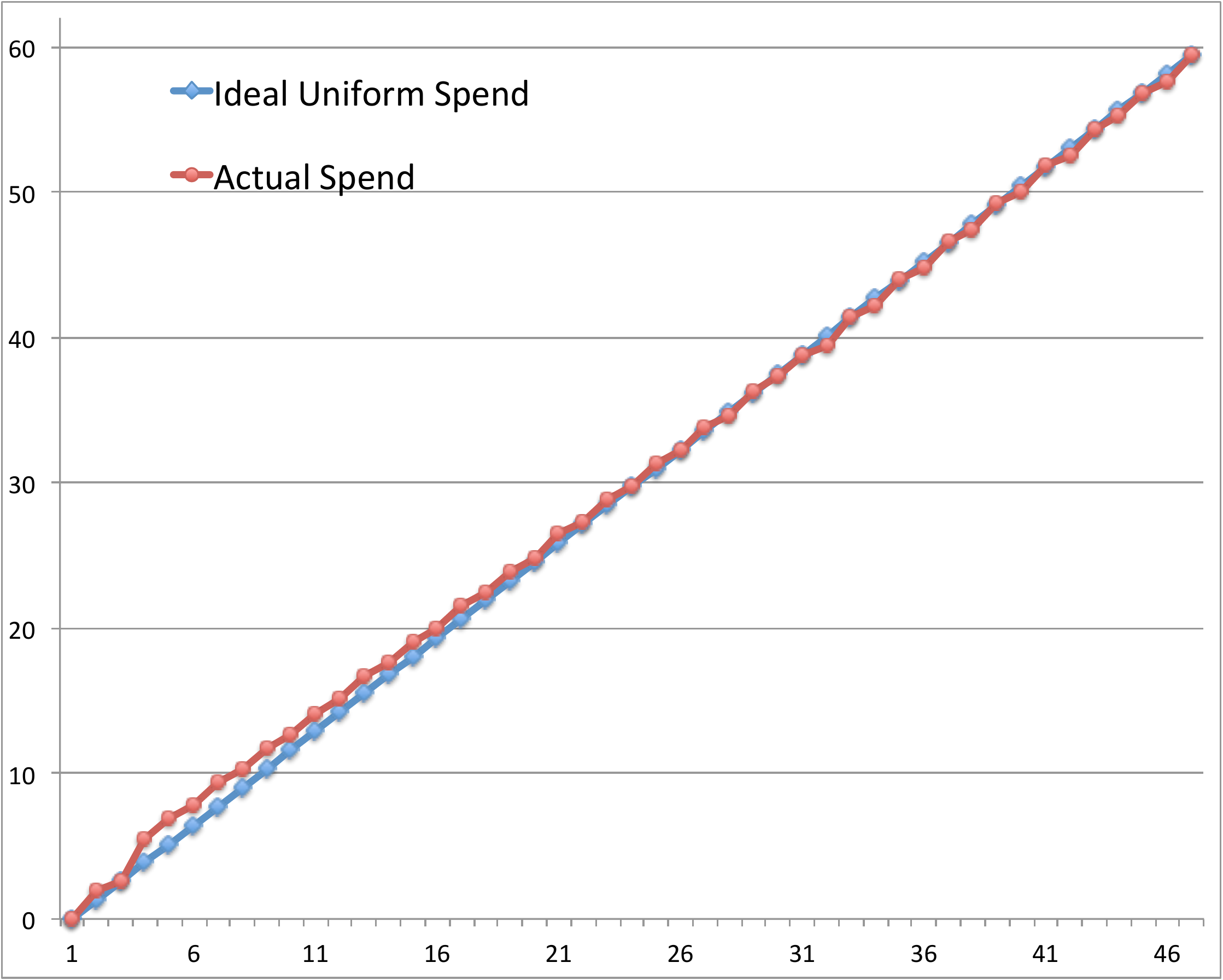}
  \caption{Simulation result of uniform budget spend. The x-axis
    represents the time slot, and the y-axis represents the total
    budget spend. The blue line depicts the ideal uniform spend, and
    the red line depicts the actual spend in our uniform budget
    pacing.}
   \label{fig:uniformpacing}
\end{figure}

\begin{figure}[t]
  \centering
  \includegraphics[width=3in]{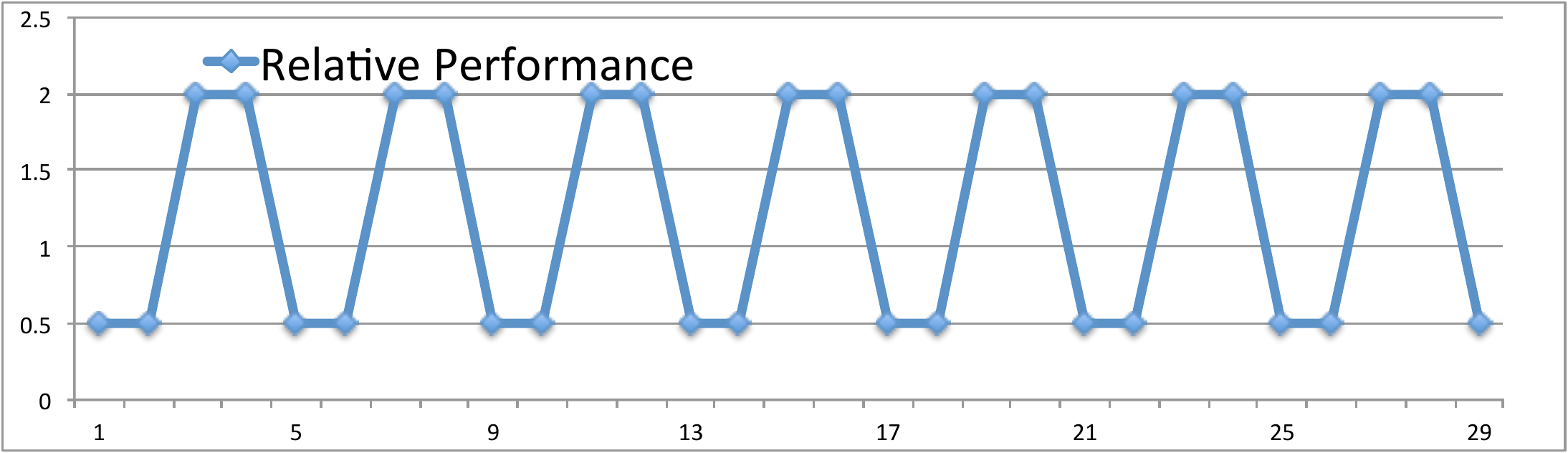}
  \caption{Relative performance across different time slot. The x-axis represents the time slot, and the y-axis represents the relative performance in terms of $\frac{1}{eCPA}$}
   \label{fig:perfdist}
\end{figure}

\begin{figure}[t]
  \centering
  \includegraphics[width=3in]{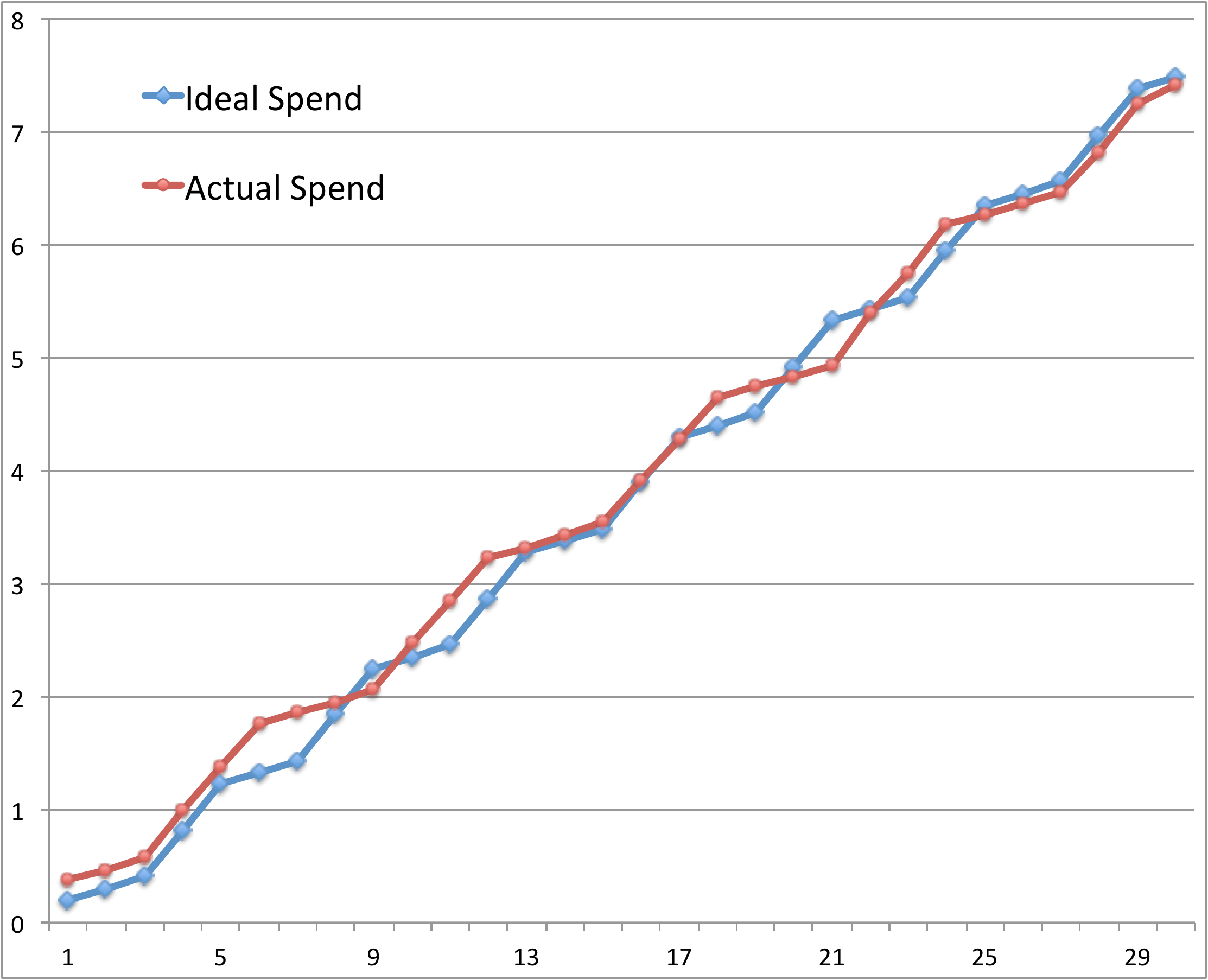}
  \caption{Simulation result of budget spend based on performance
    distribution. The x-axis represents the time slot, and the y-axis
    represents the total budget spend. The blue line depicts the ideal
    spend based on the performance distribution, and the red line
    depicts the actual spend in our performance-based budget pacing.}
   \label{fig:performancepacing}
\end{figure}

In this section, we would like to show the simulation results of the
budget pacing in our staging environment to verify that our proposed
bid optimization framework does not violate the budget constraint
specified in Eq.~\ref{eq:objective}, i.e., do not overpace or
underpace. For the simulation experiment, we launch a flat CPM
pseudo-campaign and assign a fixed amount of daily budget. A set of ad
requests are randomly generated in each time slot. The simulation
server then generates the bids based on the pacing rate and logs the
winning impressions into the database.

Fig.~\ref{fig:uniformpacing} shows the ideal uniform spend and the
actual spend using our uniform pacing strategy in
Eq.~\ref{eq:UniformSimple}. We can notice that the two lines are
pretty close to each other, and the average difference across all time
slots is $0.56$ dollars, which is less than $1\%$ error compared to
the daily budget. Fig.~\ref{fig:performancepacing} shows the ideal
spend and the actual spend using performance based pacing strategy in
Eq.~\ref{eq:PerformanceSimple} based on the relative performance shown
in Fig.~\ref{fig:perfdist}. We can notice that the actual spending
curve indeed follows the ideal spending curve, and the average
difference is $0.17$ dollars, which is about $2.3\%$ error of the
daily budget.

\subsection{Evaluation of Real Campaign Performance}

\begin{table*}[t]
\begin{center}
\begin{tabular}{|c|c|c|c|c|c|c|c|c|c|c|}
\hline
Method & DC1a & DC2a & DC3a & DC4a & DC5a & DC6a & DC7a & DC8a & DC9a & DC10a \\
\hline
Our proposal & \$1.32 & \$1.29 & \$7.92 & \$1.30 & \$1.98 & \$0.22 & \$2.77 & \$0.58 & \$3.23 & \$1.18\\
\hline
Baseline &  \$1.21 & \$1.12 & \$5.61 & \$0.8 & \$1.8 & \$0.21 & \$2.15 & \$0.53 & \$3.07 & \$1.16\\
\hline
Improvement (\%) & 9.4\% & 14.73\% & 41.14\% & 62.89\% & 9.95\% & 6.31\% & 28.37\% & 10.82\% & 5.41\% & 1.14\%\\
\hline
\end{tabular}
\end{center}
\vspace{-0.2in} \caption{CPC improvement for ten selected dynamic CPM campaigns.} 
\label{Table:dCPMCPCCampaign}
\end{table*}

\begin{table*}[t]
\begin{center}
\begin{tabular}{|c|c|c|c|c|c|c|c|c|c|c|}
\hline
Method & DC1 & DC2 & DC3 & DC4 & DC5 & DC6 & DC7 & DC8 & DC9 & DC10 \\
\hline
Our proposal & \$148.03 & \$3.14 & \$1.37 & \$59.32 & \$68.70 & \$186.55 & \$16.64 & \$3.76 & \$115.72 & \$200.27\\
\hline
Baseline &  \$206.39 & \$3.27 & \$1.34 & \$65.44 & \$72.95 & \$346.69 & \$22.93 & \$4.31 & \$204.62 & \$271.40\\
\hline
Improvement (\%) & 39.43\% & 4.45\% & -2.23\% & 10.31\% & 6.2\% & 85.84\% & 37.78\% & 14.6\% & 76.82\% & 35.51\%\\
\hline
\end{tabular}
\end{center}
\vspace{-0.2in} \caption{CPA improvement for ten selected dynamic CPM campaigns.} 
\label{Table:dCPMCPACampaign}
\end{table*}

In this section, we evaluate the entire bid optimization framework
with respect to two major classes of campaigns in our system: flat CPM
campaigns and dynamic CPM campaigns. For the evaluation of flat CPM
campaigns, the CTR metric is used and the higher rate represents
better performance. For the evaluation dynamic CPM campaigns, CPC and
CPA metrics are used based because these metrics take both total cost
of impressions and the total number of clicks and conversions into
account. The lower values for CPC and CPA metrics represent better
performance.

We first report the performance improvement in seven active flat CPM
campaigns randomly selected across different advertiser
categories. Those seven campaigns were set to run based on our
proposed approach and the existing baseline method. Each method was
run for one week and finally two weeks of data were collected for
performance comparison. Our baseline method is a simple adaptive
feedback control algorithm that multiplies a constant factor to the
current threshold of CTR $\tau(t)$ based on the pacing rate in the
time slot $t$. The first two rows shown in Table~\ref{Table:FlatCPM}
represent the CTR performance in our proposed approach and the
baseline method respectively. The third row shows the percentage of
improvement for each individual campaign. The average performance lift
achieved by our proposed approach is $123\%$.

Next we would like to report the performance improvement for dynamic
CPM campaigns. In this evaluation, we try to compare our proposed
approach with the existing baseline method that only applies the
pacing rate to uniformly select incoming ad requests without further
adjustment of bid price. Two different sets of campaigns based on the
goal type (CPC and CPA) were randomly selected across different
advertiser categories. The CPC and CPA values and the percentage of
performance lift for each individual campaign are shown In
Table~\ref{Table:dCPMCPCCampaign} and
Table~\ref{Table:dCPMCPACampaign}. We can observe that all twenty
selected campaigns running on our proposed framework perform much
better in terms of CPC and CPA metrics and the average performance
lift is $19.02\%$ for CPC campaigns, and $30.87\%$ for CPA campaigns.

\section{Conclusions}
\label{sc:conclusion}

We have presented a general and straightforward approach to perform
budget and smooth delivery constrained bid optimization for
advertising campaigns in real time. Due to the simplicity of our
algorithm, our current implementation can handle up to a million of ad
requests per second and we think it can scale to even more. Our
experimental evaluation with simulated and real campaigns shows that
our proposed algorithm provides consistent improvements in standard
performance metrics of CPC and CPA without underpacing or
overpacing. In the future, we would like to integrate the capability
of real time analytics to perform online optimization across more
user, publisher, and advertiser's attributes.

\section*{Acknowledgments}
We would like to thank Xi Yang and Changgull Song for testing the
entire bid optimization framework in the staging environment.

\bibliographystyle{natbib} 
\bibliography{kdd2013}
\end{document}